# Electronic transport properties of Co cluster-decorated graphene[*]

Chaoyi Cai(蔡超逸)[1], Jian-Hao Chen(陈剑豪) [1,2†]

[1] *International Center for Quantum Materials, Peking University, Beijing 100871*
[2] *Collaborative Innovation Center of Quantum Matter, Beijing 100871*

Interactions of magnetic elements with graphene may lead to various electronic states that have potential applications. We report an *in-situ* experiment in which the quantum transport properties of graphene are measured with increasing cobalt coverage in continuous ultra-high vacuum environment. The results show that e-beam deposited cobalt forms clusters on the surface of graphene, even at low sample temperatures. Scattering of charge carriers by the absorbed cobalt clusters results in the disappearance of the Shubnikov-de Haas (SdH) oscillations and the appearance of negative magnetoresistance (MR) which shows no sign of saturation up to an applied magnetic field of 9 T. We propose that these observations could originate from quantum interference driven by cobalt disorder and can be explained by the weak localization theory.

**Keywords:** *in-situ* quantum transport, negative magnetoresistance, weak localization

## 1. Introduction

The search for magnetic interactions in two-dimensional crystals has attracted significant attention.[1, 2] The co-existence of magnetic and electronic properties in two-dimensional (2D) crystals and heterostructures could have wide-range of applications in technologies such as data storage and computation.[3] In particular, the properties of hybrid structures consisting of non-magnetic 2D crystals and magnetic atoms are at the center of the research effort due to the potential tunability of such hybrid structures. However, interference from the environment makes it difficult to study the properties of such hybrid structures in an *ex-situ* experiment. *In-situ* atomic engineering is a powerful tool to investigate the physical properties of 2D materials and their hybrid structures.[4-6]

Graphene, a single atomic layer of graphitic carbon, is a very promising platform to explore magnetism in two dimensions owing to its extraordinary physical properties including gate-tunable carrier concentration, high electronic mobility, and versatile interactions with absorbed species.[7-12] Although many previous studies proposed utilizing defects[13, 14], non *in-situ* atomic treatment[15-17], or proximity effect[18-20] to design graphene-based magnetic devices; clear experimental evidence of magnetism in a graphene device remains elusive. A previous theory[21] predicted that among the traditional ferromagnetic elements (Fe, Co, and Ni), only Co atoms can induce magnetism in graphene. However, although several Scanning Tunneling

[*] Project supported by the National Basic Research Program of China under Grant Nos 2013CB921900 and 2014CB920900, the National Natural Science Foundation of China under Grant No 11374021).
[†] Corresponding author. E-mail: chenjianhao@pku.edu.cn

Microscopy (STM) experiments[22-24] for Co-decorated graphene on metal substrates have been reported, there exists no transport experiment for Co-decorated isolated graphene on an insulating substrate to show the presence of induced magnetism.

In this work, we performed *in-situ* quantum transport measurement to study the effect of the electronic properties of Co-decorated graphene. The results show that when the cobalt atoms are deposited on the cold surface of graphene, clusters are naturally formed on the surface of graphene, which introduce both long-range and short-range scatterings. Negative magnetoresistance (MR) that does not saturate up to a magnetic field of 9 T has been observed. No transport signature of ferromagnetism has been found, although the possibility of induced paramagnetism cannot be excluded. We discuss the possible reasons for the absence of transport signature of ferromagnetism and the origin of the negative MR by comparing our results with the relevant theories and experiments. In addition, a metal-insulator transition driven by quantum interference is observed and may be related to the negative MR. These observations cannot be explained by the spin-half paramagnetism theory, variable range hopping (VRH) model, and the opening of an energy gap, but can be possibly explained by weak localization.

## 2. Experimental procedures

Graphene was mechanically exfoliated from a flake of Kish graphite onto 300 nm $SiO_2$/Si substrates. Standard electron-beam lithography and metallization processes were used to make Hall bar structures. Electrodes were made of 5 nm Cr/50 nm Au and a four-wire configuration was used in the measurements. The device was then transferred into our homemade *in-situ* quantum transport apparatus. The system consists of an ultra-high vacuum (UHV)-compatible, Gifford-McMahon (GM) cryocooler-based cryostat from Advanced Research Systems Co. Ltd., a dry 9 T solenoid magnet from Cryogenic Co. Ltd., and a special UHV chamber with surface modification apparatus. An e-beam evaporator was used in this experiment to produce high purity cobalt flux. Seven runs of cobalt deposition were conducted and the transport properties of graphene after each run were studied *in-situ*. Continuous UHV environment was maintained during the whole process of deposition and measurement.

## 3. Results and Discussion

Figure 1 shows the impact of cobalt adsorption on the conductivity of graphene sheets. Changes induced by cobalt adsorption are the following: (1) A decrease of conductivity and mobility; (2) a shift in gate voltage of minimum conductivity ($V_{min}$) to more negative gate voltage; (3) a decrease in minimum conductivity, $\sigma_{min}$; (4) an additional gate-dependent resistivity, which varies as $|V_g - V_{min}|^\delta$. Features (1), (2), and (3) indicate that cobalt adsorption generates n-type doping and is consistent with charged impurity scattering.[4, 5] Figure 1(b) shows additional resistivity due to cobalt adsorption during different rounds as a function of $|V_g - V_{min}|$. The nonlinearity of additional resistivity with $|V_g - V_{min}|$ shown in the inset of figure 1(b) can be attributed to incomplete screening of the potential imposed by cobalt on graphene, similar to previous atom-doping experiments[25, 26]. However, feature (4) differs from point-like charged impurities, as we will discuss in more detail subsequently.

Figure 2 is obtained by plotting data of Figure 1(b) on a logarithmic scale. As shown in Figure 2, the additional resistivity as a function of gate voltage was carefully compared between electron and hole carriers for different runs. The asymmetry of additional resistivity between the electron side and the hole side decreases with increasing cobalt doping. The observation indicates that cobalt adsorption introduced scattering that differs from scattering of pure charge impurities.[4, 27] In addition, the additional resistivity varies as $n^\delta$, where $n$ is the carrier density $\propto |V_g - V_{min}|$ and $\delta$ is the exponent. We found that with increasing doping concentration at high carrier densities (Figure 3(a)), $\delta$ saturates at 0.8. Previous theories point out that the density dependent resistivity of graphene is expected to be $\rho \propto n^\delta$, with $|\delta| = 1$ for charged impurities and $|\delta| < 1$ for short-range scattering.[4, 28] These observations show that the cobalt adsorption will not only introduce long-range scattering, but also introduce short-range scattering in graphene.

As scattering from single ions and clusters of charged impurities show a similar resistivity dependence on carrier density, it is important to investigate the relationship between the mobility ($\mu$) and the shift in Dirac points ($V_{shift} = |V_{g,min} - V_{0,min}|$) to determine whether these features described above are generated by scattering from single ions or from clusters of charged impurities. The addition of charged impurities will result in a shift of the Dirac point in graphene through the power-law relationship, $V_{shift} \sim (1/\mu - 1/\mu_0)^b$, where $\mu_0$ is the mobility of the sample before Co-decoration. For random point-like charged impurities, each impurity contributes equally to doping and in providing scattering cross section, leading to $V_{shift} \propto 1/\mu$; however, for clusters of charged impurities, each impurity atom contributes equally to doping in a similar manner as in the case of random point-like impurities, but the scattering cross section per atom is significantly reduced and dependent on cluster size, leading to $V_{shift} \propto (1/\mu)^b$, where $b < 1$.[26, 29] The slope of the graph obtained by plotting the experimentally obtained $V_{shift}$ versus $1/\mu$ in a semi-log plot provides the value of the exponent $b$. Figure 3(b) shows our data as well as power-law fitting with $b = 0.81$ for electrons and $b = 0.58$ for holes. The experimentally determined coefficient $b$ indicates that cobalt deposited by an e-beam evaporator forms clusters on the graphene surface, even though the graphene sample was held at cryogenic temperatures.

Next, the temperature dependence of the resistivity at zero magnetic field was studied. It is well known that pristine graphene is highly conductive with a weakly temperature-dependent resistivity as shown in the inset of Figure 4(a). After cobalt decoration, the resistivity becomes more sensitive to temperature as shown in Figure 4(a). The device behaves as an insulator, especially at low temperature and at the vicinity of the Dirac point (figure 4(b)). Previous studies reported that disorder in graphene obtained by hydrogenation or oxidation could open an energy gap in graphene.[30-32] We plotted $\rho_{Dirac}$ as a function of inverse temperature (inset of Figure 4(b)) on logarithmic scale and found that $\rho_{Dirac}$ was poorly described by thermal activation model $\rho \propto e^{E_g/k_B T}$. The energy gap obtained from the fitting was $E_g \sim 0.1$ meV, which is much smaller than both the lowest achieved sample temperature $k_B T \sim 3$ meV and the smallest achieved potential fluctuations due to charge inhomogeneity (of the order of 98 meV near the Dirac point for the pristine sample).[33] Thus the possibility of the formation of energy gap as the mechanism for the insulating behavior of cobalt-decorated graphene can be excluded.

We also note that even for the highest cobalt concentration, the resistivity of graphene at the Dirac point remains smaller than $h/e^2$ at the lowest temperature of 7.1 K, increasing only twice compared to the resistivity of the Co-decorated sample at 300 K. Therefore, our data cannot be explained by 2D variable-range hopping (2D VRH). Besides, 2D VRH would be accompanied by a positive MR due to the shrinkage of electron wavefunction when the sample is subject to increasing perpendicular magnetic field,[34-36] which counters our experimental observation (shown in the next paragraph).

Now we focus our discussion on the transport properties of Co-decorated graphene under perpendicular magnetic field, in which non-saturating negative MR was observed. Figure 5(a) shows that the magnitude of the negative MR increases as the density is lowered. Figure 5(b) plots the dependence of MR on cobalt concentration for different doping runs. The disappearance of SdH oscillation with increasing doping can be observed and the negative MR becomes more pronounced without signs of saturation up to the highest magnetic field of 9 T achieved in this experiment. In existing theories, the negative MR in perpendicular magnetic field originates from the alignment of magnetic moments in the sample due to the external magnetic field or from the destruction of localization driven by quantum interference. If the negative MR is attributed to ferromagnetism introduced by cobalt, anomalous Hall resistivity in $R_{xy}$ vs. $B$ and hysteresis in $R_{xx}$ vs. $B$ as well as $R_{xy}$ vs. $B$ when cycling magnetic field between $B_{max}$ and $-B_{max}$ should be observed below the Curie temperature. However, no such signatures are found which indicates that the device is nonmagnetic or paramagnetic. Previous studies[16] reveal that point defects will introduce notable paramagnetism to the graphene sheet. In paramagnetic materials, the global magnetization $M$ should behave like the Brillouin function $M = NgJ\mu_B \left[ \frac{2J+1}{2J} ctnh\left(\frac{(2J+1)z}{2J}\right) - \frac{1}{2J} ctnh(\frac{z}{2J}) \right]$, where $z = \frac{gJ\mu_B H}{k_B T}$, $g$ is the $g$-factor and $J$ is the angular momentum. The relationship between $MR$ and $M$ in general is $MR \propto -M^2(H)$ in multicarrier systems.[37] If such relationship could be applied to graphene in the vicinity of the Dirac point, we could extract $M(H)$ from the MR data of Figure 5. However, our MR data does not fit to the Brillouin function. A reasonable speculation is that cobalt clusters are different from point defects, which create $P_z$ vacancies that are expected to carry magnetic moments. Hence, we deduce that magnetism did not play an important role in the origin of the negative MR in our sample. In the following, another possible explanation for the negative MR, the suppression of quantum interference by increasing perpendicular magnetic field, is discussed.

In disordered systems with sufficiently long electron dephasing length, time-reversal symmetric scattering paths could constructively interfere, leading to a logarithmic divergence in the resistivity of the sample with decreasing temperature,[38] an effect that is much more pronounced for 2D systems.[39, 40] The Co-decorated graphene indeed shows logarithmic divergence in resistivity as shown in Figure 4(b), consistent with the appearance of weak localization in the sample. In systems with weak localization, application of the magnetic field perpendicular to the sample breaks the time-reversal symmetry between forward and backward hopping paths, destroying the quantum inferences thus generating a negative MR. The correction of conductivity to the semi-classical Drude conductivity in this case is given by:[41]

$$\Delta\sigma = \frac{e^2}{\pi h}\left[F\left(\frac{B}{B_\varphi}\right) - F\left(\frac{B}{B_\varphi+2B_i}\right) - 2F\left(\frac{B}{B_\varphi+B_i+B_*}\right)\right]. \quad (1)$$

Here, $F(x) = \ln(x) + \psi\left(\frac{1}{2}+\frac{1}{x}\right)$, $B_{\varphi,i,*} = \frac{\hbar}{4eL_{\varphi,i,*}}$, and $\psi(x)$ is the digamma function. We explore whether the negative MR results from time-reversal symmetry breaking, the phase coherence length, $L_\varphi$, and inter-valley scattering length, $L_i$, are the necessary fitting parameters. By tuning the back gate, the carrier density $n$ of graphene can be varied from $1.4 \times 10^{16}\,m^{-2}$ to $4.2 \times 10^{16}\,m^{-2}$ after each round of cobalt deposition. Then the low-field magneto-conductance (MC) for different runs was compared for the same carrier density. Figure 6 shows the plot of the experimental MC data, which agrees well with the theory. The phase coherence length $L_\varphi$ and inter-valley scattering length $L_i$ can be extracted from Figure 6. Figure 7(a) shows dependence of $L_\varphi$ and $L_i$ on carrier density. The increase of $L_\varphi$ with increasing $n$ and the weak dependence of $L_i$ on $n$ is consistent with previous works.[42, 43] The more convincing evidence comes from Figure 7(b) where both length scales decrease with increasing cobalt coverage. The same trend of $L_\varphi$ and $L_i$ indicates the correlation between cobalt adsorption and quantum interference driven by short-range scattering centers, consistent with our finding in Figure 2. The order of magnitude of $L_\varphi$ is in good agreement with the previous results in disordered graphene.[44]

To understand the fact that $L_\varphi$ and $L_i$ both decrease with increasing cobalt decoration and decreasing carrier concentration, one can make use of the concept of localization length without involving magnetic interactions. Even though a quantitative theory describing the localization length, $\xi$, in a weakly localized sample (e.g. Co-decorated graphene) is yet to be developed, $\xi$ should be decreasing in a sample with increasing disorder (high cobalt concentration) and decreasing screening (lower carrier density). On the other hand, $L_\varphi$ and $L_i$ should be bounded by $\xi$ by definition. Thus, the trends shown in Figure 7 are easily understood and a direct correlation between the increasing magnitudes of the negative MR in Co-decorated graphene and a decreasing $\xi$ can be established.

One interesting finding in this experiment is that there is no sign of saturation of the negative MR even at a magnetic field of 9 T, which is relatively large for weak localization. A phenomenological explanation can be used to understand this. In the weak localization picture, the comparison between two length scales is critical. The first length is the phase coherence length, $L_\varphi$, and the other is the magnetic length, $l_B = (\hbar/eB)^{1/2}$, which is the cyclotron radius of the electrons. The suppression of quantum interference in disordered systems happens approximately when $L_\varphi > 2\pi l_B$, and when the resistivity of the sample recovers its classical value.[45, 46] Since the upper bound of $2\pi l_B = 2\pi(\hbar/eB)^{\frac{1}{2}} = 53.8$ nm at $B = 9$ T, which is still larger than $L_\varphi$, the appearance of the negative MR in the whole range of measurement is reasonable.

## 4. Conclusion

We have systematically studied the transport properties of cobalt-decorated graphene by *in-situ* quantum transport measurements. We found that e-beam-evaporated cobalt formed clusters even on the cold surface of graphene and introduced both long-range and short-range scatterings. The Shubnikov-de Haas (SdH) oscillations disappeared with increasing cobalt coverage and no signature of ferromagnetism was been observed. Cobalt-decorated graphene exhibited a negative MR in perpendicular magnetic field and an insulating behavior with decreasing temperature. The above behavior can be explained well in the framework of weak localization caused by cobalt adsorption.

**Figure Captions**

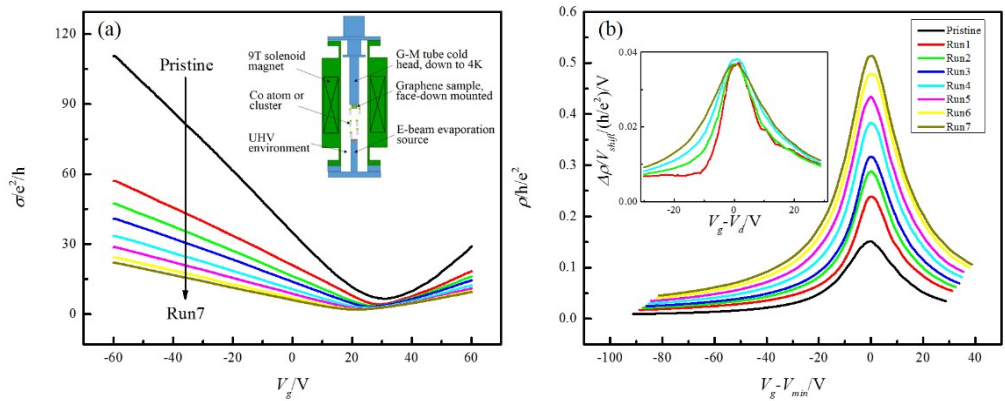

**Figure.1.** (a) The conductivity (σ) versus gate voltage ($V_g$) curves for the pristine sample and seven different doping concentrations obtained at $7.1\ K$ in UHV. Inset: Schematics of the experimental setup. (b) Resistivity as a function of $V_g - V_{min}$ at different doping concentrations. Inset: Added

resistivity as a function of $V_g - V_{min}$ at different areal dosage normalized to $V_{shift} = |V_{g,min} - V_{0,min}|$. The color code is the same for the inset and the main graph of Figure 1(b).

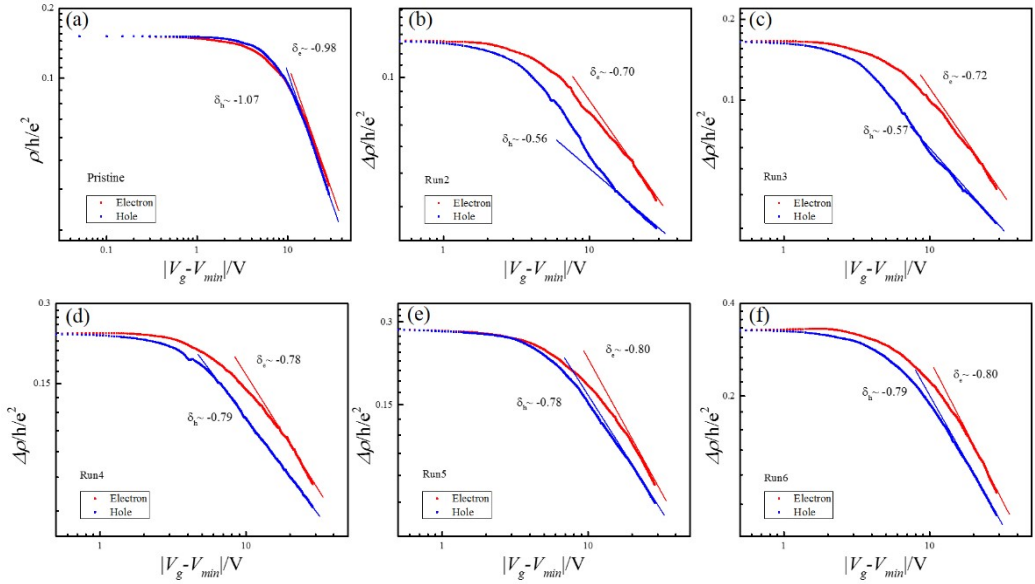

**Figure.2.** (a) Log-log plot of gate dependence of the resistivity as a function of $|V_g-V_{min}|$ for the pristine sample. (b)-(f) Log-log plot of gate dependence of the added resistivity as a function of $|V_g-V_{min}|$ for Run 2-6. The blue and red dots represent electrons and holes respectively. The solid line indicates the fitting to $|V_g - V_{min}|^\delta$.

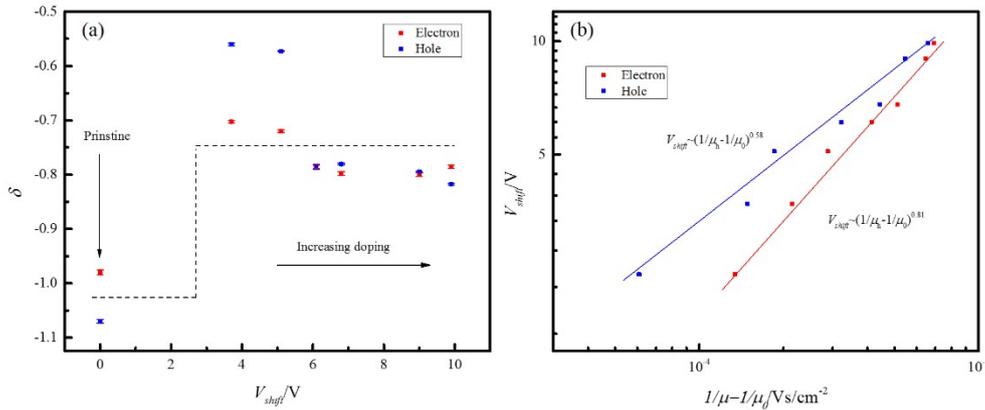

**Figure.3.** (a) The exponent $\delta$ vs. $V_{shi}$ which represents the shift of the Dirac point due to Co-absorption. The dashed line indicates jump of the exponents between the pristine resistivity data and the added resistivity from Co-absorption. (b) $V_{shift}$ vs. $(1/\mu - 1/\mu_0)$. The solid line is the power-law fit to the equation, $V_{shift} \sim (1/\mu - 1/\mu_0)^b$.

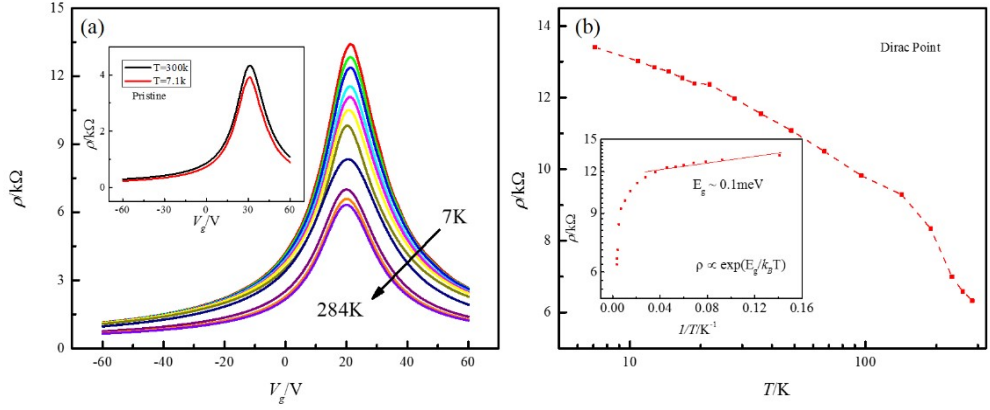

**Figure.4.** (a) Gate dependence of the resistivity at different temperatures for Run 7. Temperature is labeled using different colors. Inset: Gate dependence of the resistivity of pristine graphene at $T = 7.1$ K and $T = 300$ K. (b) Temperature dependence of the resistivity at Dirac point for Run 7. Inset: $\rho_{Dirac}$ vs $1/T$ in a semi-log plot. The solid line is the thermal activation model fitting as described in the main text.

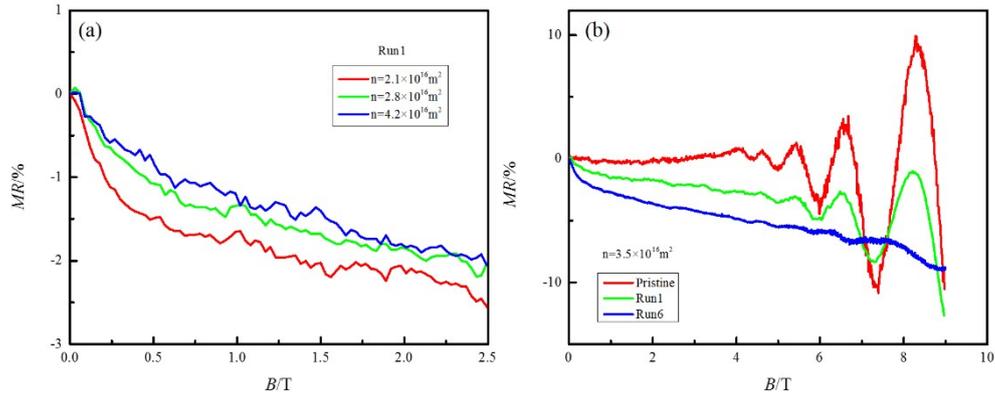

**Figure.5.** (a) MR in the perpendicular magnetic field for Run 1 at $T = 7.1$ K and three different carrier densities $n = 2.1,\ 2.8,$ and $4.2 \times 10^{16} \mathrm{m}^{-2}$. (b) MR for carrier density $n = 3.5 \times 10^{16} \mathrm{m}^{-2}$ for pristine graphene, Run 1 and Run 6.

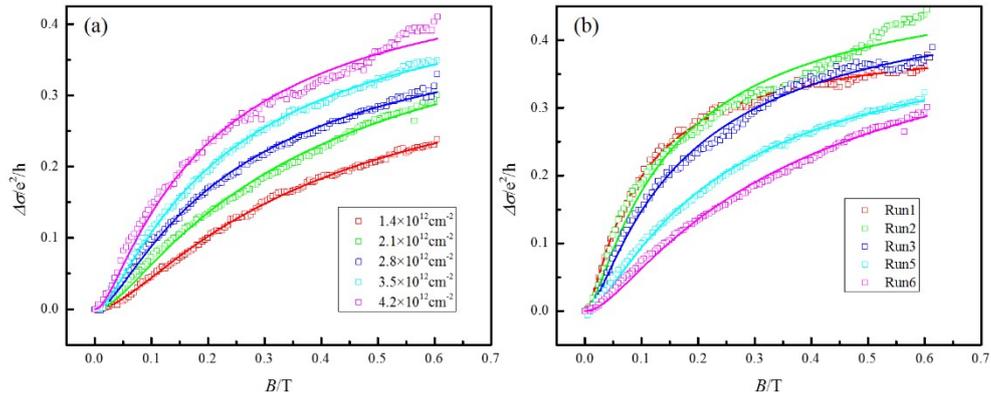

**Figure.6.** (a) Magnetoconductance as a function of magnetic field for carrier density varied from $n = 2.1$ to $4.2 \times 10^{12} \text{cm}^{-2}$ for Run 6. (b) Magnetoconductance as a function of magnetic field for carrier density $n = 2.1 \times 10^{12} \text{cm}^{-2}$ during different Runs. Fits to Eq. (1) are shown by solid lines.

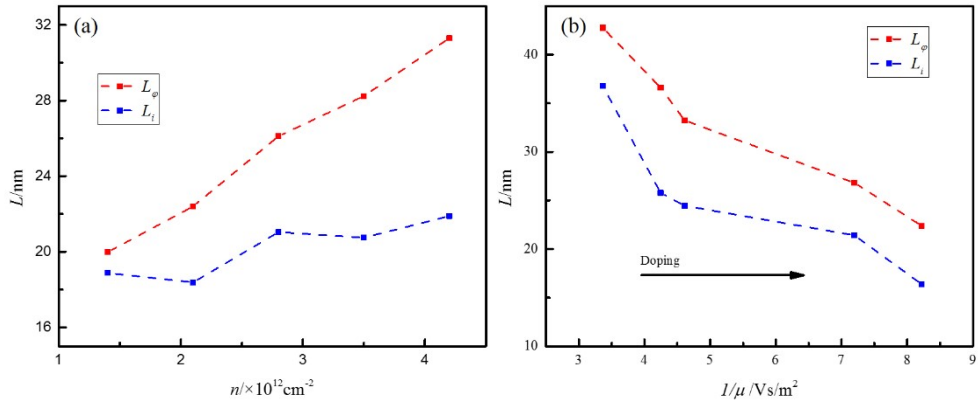

**Figure.7.** Characteristic length as a function of (a) carrier density and (b) $1/\mu$, respectively. Dashed lines are guides to the eyes.